\documentclass[lettersize,journal]{IEEEtran}
\usepackage{amsmath,amsfonts}
\usepackage{algorithmic}
\usepackage{algorithm}
\usepackage{array}
\usepackage[caption=false,font=normalsize,labelfont=sf,textfont=sf]{subfig}
\usepackage{textcomp}
\usepackage{stfloats}
\usepackage{url}
\usepackage{verbatim}
\usepackage{graphicx}
\usepackage{cite}
\usepackage{hyperref}
\usepackage{booktabs}

\usepackage{enumitem}
\usepackage{flushend}
\usepackage{xcolor}
\usepackage{hyperref}

\newcommand{\name}{\textsc{GNN101}}

\newcommand{\qw}[1]{\textcolor{black}{#1}}
\newcommand{\harry}[1]{\textcolor{black}{#1}}
\newcommand{\chongwei}[1]{\textcolor{black}{#1}}

\newcommand{\ie}{\textit{i.e.}}
\newcommand{\eg}{\textit{e.g.}}
\newcommand{\etal}{\textit{et al.}}
\usepackage{cleveref}

\usepackage{ulem}

\begin{document}

\title{GNN101: Visual Learning of Graph Neural Networks in Your Web Browser}

\author{Yilin Lu, Chongwei Chen, Yuxin Chen, Kexin Huang, Marinka Zitnik, Qianwen Wang
\thanks{This paper was produced by the IEEE Publication Technology Group. They are in Piscataway, NJ.}
\thanks{Manuscript received April 19, 2021; revised August 16, 2021.}}

\markboth{Journal of \LaTeX\ Class Files,~Vol.~14, No.~8, August~2021}%
{Shell \MakeLowercase{\textit{et al.}}: A Sample Article Using IEEEtran.cls for IEEE Journals}

\IEEEpubid{0000--0000/00\$00.00~\copyright~2021 IEEE}

\maketitle

\begin{abstract}
Graph Neural Networks (GNNs) have achieved significant success across various applications. However, their complex structures and inner workings can be challenging for non-AI experts to understand. 
To address this issue, this study presents \name{}, an educational visualization tool for interactive learning of GNNs. 
\name{} introduces a set of animated visualizations that seamlessly integrates mathematical formulas with visualizations via multiple levels of abstraction, including a model overview, layer operations, and detailed calculations. 
Users can easily switch between two complementary views: a node-link view that offers an intuitive understanding of the graph data, and a matrix view that provides a space-efficient and comprehensive overview of all features and their transformations across layers.
\name{} was designed and developed based on close collaboration with four GNN experts and deployment in three GNN-related courses.
We demonstrated the usability and effectiveness of \name{} via use cases and user studies with both GNN teaching assistants and students.
To ensure broad educational access, \name{} is developed through modern web technologies and available directly in web browsers without requiring any installations.

\end{abstract}


\section{Introduction}

\IEEEPARstart{G}{raph} Neural Networks (GNNs) offer powerful capabilities for analyzing graph-structured data (\eg, social networks, molecular graphs).
This type of data is often difficult to be effectively modeled by traditional machine learning (ML) models, which are mainly designed for non-graph data such as images and text.
Therefore, GNNs have earned increasing popularity, especially in AI4Science research that leverages the inherent graph structures in scientific data to drive discoveries (\eg, new material design and drug development)~\cite{huang2024foundation, wang2022extending}.

However, learning GNNs can be more challenging than learning ML models for non-graph data, due to the unique characteristics of graphs and the complexities of graph-based computations. Unlike structured data such as images or tables, graphs are inherently non-Euclidean, meaning their data representation cannot be neatly organized into fixed-size grids. This irregularity makes it difficult to interpret standard computational operations like convolutions or pooling.
Meanwhile, graphs can be sparse or have massive sizes, posing significant computational and memory challenges that require special techniques like graph sampling, mini-batching, and message passing.
These data processes introduce additional complexity for learning GNNs.

\IEEEpubidadjcol

Recently, a wide range of GNN educational resources has emerged, presented in various formats \harry{such as online blogs ~\cite{datacamp_gnn_tutorial, sanchez-lengeling2021gentle, daigavane2021convolution}, lecture videos~\cite{raschka2020gnn, standford224w, uvadlc}, and computational notebooks~\cite{uvadlc2021gnn, TF-GNN, pyg-official}}.  While these resources offer significant value, they often rely heavily on static illustrations (\eg, diagrams or mathematical equations) to explain GNNs~\cite{standford224w, upennGNN}. 
As a result, they tend to either focus on high-level concepts with limited attention to the detailed inner workings of GNNs, or concentrate on implementation details in specific programming frameworks. 
This creates a gap in bridging theoretical concepts with detailed computations via intuitive and interactive approaches.

\begin{figure}
  \centering
  \includegraphics[width=0.9\linewidth, alt={interface of gnn where main features are highlighted.}]{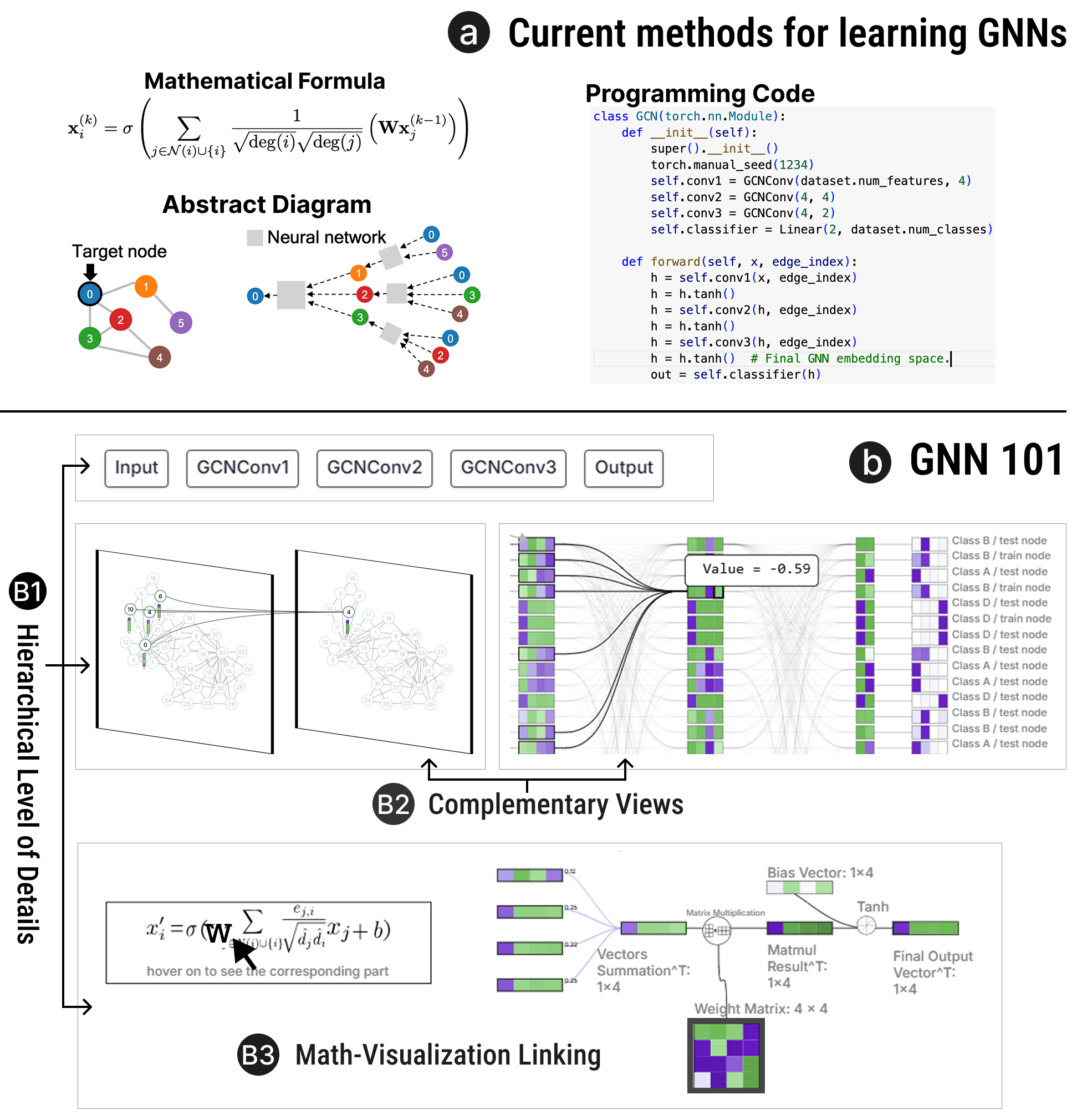}
  \caption{%
    (a) Traditional methods for learning GNNs.
    (b) \name\ enhances understanding of GNNs through a hierarchical breakdown of details (B1), complementary views (B2), and the integration of mathematical concepts with visualizations (B3).
  }
  \label{fig:teaser}
\end{figure}

Interactive visualization has long been recognized as an effective method for understanding algorithms (\eg, sorting, searching~\cite{hundhausen2002meta}) and ML models (\eg, convolution neural networks~\cite{wang2020cnnexp}, multi-layer perceptron~\cite{neuralPlayGround}).
By making the intricate processes and concepts of MLs more accessible and intuitive, interactive visualization offers a promising approach to facilitate more effective learning of GNNs. 
However, visualizing GNNs can be challenging due to the inherent complexity of graph structures and the extensive mathematical details involved in GNN computations.
While recent research has proposed visualization systems for GNNs, such as GNNLens~\cite{jin2022gnnlens} and CorGIE~\cite{liu2022visualizing}, these tools are primarily aimed at AI developers for debugging and improving models.
They focus on specific aspects related to model performances, such as error patterns~\cite{jin2022gnnlens} and embedding qualities~\cite{liu2022visualizing}, and provide limited support for beginners to learn about GNNs.
Most relevant to this paper are the two interactive articles about GNNs from Distill \cite{sanchez-lengeling2021gentle, daigavane2021convolution}, which effectively use interactive visualizations to explain various core GNN concepts (\eg, adjacency matrix, convolution, pooling). 
However, the two interactive articles primarily focus on isolated concepts, lacking a cohesive connection within the context of a complete GNN model. 
Additionally, these articles mainly provide abstract examples for users to interact with, such as an abstract graph with four nodes (A, B, C, D) where each node is represented by a single numerical feature. 
Consequently, they fall short of bridging the gap between theoretical understanding and practical execution process of real-world GNN predictions.

To address these limitations and provide a more effective learning experience for GNNs, we introduce \textbf{\name, the first interactive visualization tool for learning GNNs}.
The design of \name\ is informed by challenges and requirements identified via a thorough analysis of existing GNN educational materials, and close discussions with instructors and students.
To address the challenges in educational GNN visualization, 
\name\ features a seamless linkage between mathematical formulas and visualizations via different levels of GNN details and provides two complementary views to enhance the learning experience (\autoref{fig:teaser}).
Specifically, it integrates a model overview, layer operations, and detailed animations for matrix calculations with smooth animations.
We are currently deploying \name\ in three courses that involve the teaching of GNNs at three different universities. This deployment has led to usage scenarios and observational studies that demonstrate the usability of \name\ and generate design lessons for educational AI visualization tools.
The main contribution of this paper includes:
\begin{itemize}[noitemsep, leftmargin=*]
    \item \textbf{The design and development of \name}, an interactive educational visualization tool that helps AI beginners to learn the working mechanisms of GNNs. 
    \name\ is open-source and can be viewed online without any installation.
    %
    \item \textbf{Novel interaction, visualization, and animation designs} that seamlessly link mathematical formulas and model visualizations while providing different levels of GNN model details with a modified focus + context visualization. 
    \item \textbf{Evaluations} that demonstrate the effectiveness of \name{} via in-lab user studies and real-world deployment,
    with corresponding \textbf{design implications} discussed.
\end{itemize}





\section{Related Studies}

\subsection{Algorithm Visualization}
Developing algorithm visualizations (AV) to facilitate learning has a long history and can be traced back to the 1970s~\cite{hundhausen2002meta}.
These AV tools typically represent data structures and values via graphical elements, and illustrate their changes via animated transitions to illustrate the step-by-step execution of algorithms~\cite{stasko1990tango, gurka1996testing, guo2013online, naps2005jhave}.  
Despite the promise of AV, studies found educators tend to stick to more traditional pedagogical technologies (\eg, whiteboards and overhead projectors), due to the difficulties in adopting AV techniques and the mixed results regarding its educational effectiveness~\cite{hundhausen2002meta, naps2005jhave, gurka1996testing}.
To promote AV's accessibility, researchers have explored tools that require minimal setup or customization. 
For example, online python tutor~\cite{guo2013online} employs web-based technologies to support program visualization directly in a web browser without any installation, which has contributed to its widespread usage (over 30,000 users per month).
To better understand AV's effectiveness,
Hundhausen \etal~\cite{hundhausen2002meta} conducted a systematic meta-study of 24 experimental studies. Their findings suggest that how students engage with AV technology has a greater impact on learning outcomes than the content the technology presents. 
Similarly, Byrne \etal~\cite{byrne1999evaluating} discovered that animations enhance learning by encouraging learners to predict algorithmic behavior.

\name\ builds on prior AV studies, particularly inspired by their use of animated visual transitions to enhance user engagement and their emphasis on easy accessibility.
Additionally, \name\ explores how to employ these design insights to the new context of learning GNNs. This context introduces unique challenges due to the large volume of data and the complexity of the computational processes involved.

\subsection{Educational ML Visualization}
With the rise of ML, educational visualizations of ML models have gained significant popularity. 
These visualizations are being developed for a wide range of ML models, including CNNs~\cite{liu2016cnnvis, wang2020cnnexp}, RNNs~\cite{strobelt2017lstmvis, ming2017rnnvis}, GANs~\cite{wang2018ganviz, kahng2018gan}, and transformers~\cite{yeh2023attentionviz, LLMVIS}. 
To promote user engagement, these educational visualizations often feature direct model interactions, \ie, providing real-time visualizing model intermediate states as users run predictions and train models.
For example, ConvNetJS demo~\cite{karpathy2014convnetjs}, CNN Node-Link Visualization~\cite{harley2015isvc}, CNNExplainer~\cite{wang2020cnnexp}, and LLM Visualization~\cite{LLMVIS} support real-time predictions of selected data points by running the ML model in web browser, and visualize the model internal states during the prediction.
Teachable Machine~\cite{carney2020teachable}, TensorFlow Playground~\cite{neuralPlayGround}, and GANLab~\cite{kahng2018gan} allow users to train a deep neural network classifier with data collected from their own web camera, or from the provided example datasets.
Apart from web-based tools, interactive Distill articles~\cite{olah2017feature, olah2018building} that combine text tutorials with interactive visualization are gaining popularity as an alternative medium for education.

While providing valuable educational support, these studies focus on explaining either the high-level model structures\harry{~\cite{kahng2018gan, cho2025transformer, sanchez-lengeling2021gentle}} or the low-level mathematics\harry{~\cite{daigavane2021convolution,datacamp_gnn_tutorial}}, missing effective mechanisms to connect both.
CNNExplainer~\cite{wang2020cnnexp} is the most relevant by connecting high-level model structures and low-level numerical operations, which greatly inspired the hierarchical level of details design in \name.
However, CNNExplainer is specifically designed for Euclidean data (\eg, images and text) and cannot be directly applied to GNNs.


\subsection{GNN Visualizations}
Recent research has proposed various visualization systems for GNNs, such as GNNLens~\cite{jin2022gnnlens}, CorGIE~\cite{liu2022visualizing}, GNNAnatomy~\cite{lu2024gnnanatomy}, GNNFairViz~\cite{ye2024gnnfairviz}.
These tools primarily target AI developers to assist with model debugging~\cite{jin2022gnnlens}, or support domain users in AI-assisted decision-making~\cite{wang2022extending}.
Research has shown that AI beginners and experts have significantly different requirements when it comes to visualizing ML models~\cite{yuan2021survey, wang2022extending}. As a result, these existing GNN visualization tools cannot be applied for learning purposes.

Most relevant to our study are the interactive Distill articles by Daigavane \etal~\cite{daigavane2021convolution} and Sanchez-Lengeling \etal~\cite{sanchez-lengeling2021gentle}.
These articles combine interactive GNN visualizations with text tutorials, but tend to use simplified data and visualizations. 
For example, the GNN Playground proposed by Sanchez-Lengeling shows the embedding of graphs rather than the complete layer-by-layer computation within a GNN.
Daigavane \etal only illustrated a single layer and utilized scalar numbers to represent node features, which are typically high-dimensional in real-world GNN applications.
This gap underscores the need for more comprehensive and realistic educational visualizations for GNNs.

\begin{table*}[]
    \centering
    \small
    \begin{tabular}{|p{1.1cm}|p{1.4cm}|p{1.5cm}|p{1.8cm}|p{1.6cm}|p{1.6cm}|p{1.5cm}|p{1.0cm}|p{1.0cm}|}
    \hline
    \multicolumn{9}{|c|}{\textbf{Which Concepts are Taught}} \\
    \hline
      \multicolumn{2}{|c|}{Data: 17/17}   & \multicolumn{4}{c|}{Inside a GNN Layer: 17/17} & \multicolumn{3}{c|}{Model Architecture: 17/17}  \\
      \hline
      Graph \newline Structure: 17/17 & High-Dim \newline Features: 15/17  & Layer Input and Output: 17/17   &  Aggregation of Neighbors: 14/17 &  Weights of Neighbors: 13/17 & Sampling of Neighbors: 9/17 & Non-GNN Layer: 11/17 & Different Tasks: 14/17 & GNN \newline Variants: 13/17 \\
       \hline
        \multicolumn{9}{|c|}{\textbf{How are They Taught}} \\
      \hline
         \multicolumn{2}{|c|}{Mathematical Formula: 15/17}   & \multicolumn{2}{c|}{Abstract Diagram: 14/17} & \multicolumn{2}{c|}{Python Code: 10/17} & \multicolumn{3}{c|}{Visualization of Real Data: 4/17} \\
       \hline
    \end{tabular}
    \caption{ \qw{Summary of key GNN concepts and teaching methods covered across 17 educational tutorials. 
    The top half shows coverage of key concepts in three categories, while the bottom half shows different teaching modalities.}}
    \label{tab:gnn_content}
\end{table*}

\section{Designing \name}
\label{sec:design}

To make \name\ an effective educational tool for \qw{novice AI practitioners to learn} GNNs, we need to answer two main questions: 
1) \textit{What types of information are essential for learning GNNs?} and 2) \textit{What challenges arise when learning such information?}
To answer these questions, we conducted a thorough review of 17 different GNN educational resources and engaged in close collaboration with four GNN experts.
\qw{
The four GNN experts are all experienced AI researchers, including two professors (E1, E2) who have each taught courses on GNNs for more than four years, and two GNN researchers (E3, E4) who have extensive experience in educating the GNN user community as core maintainers of popular GNN libraries. 
E1 and E3 are also co-authors of this paper.
Discussions with these experts took place during both the initial design phase and throughout the iterative feedback process.}


\subsection{Reviewing Existing GNN Tutorials}
To answer \textit{``what types of information are essential for learning GNNs''},
we analyzed 17 different GNN educational resources gathered through online searches and the recommendations from our GNN experts.
These resources include not only high-impact GNN courses~\cite{standford224w, pyg-tutorial, uvadlc}, but also interactive articles~\cite{sanchez-lengeling2021gentle, daigavane2021convolution}, official tutorials from widely-used GNN libraries~\cite{TF-GNN, pyg-official}, and highly-rated YouTube videos.
Two authors independently coded the content of these tutorials to determine \textit{which concepts were taught} and \textit{how were they taught}.
The initial open codes were then systematically organized through axial coding. The list of tutorials and their codes is available in the supplementary material.

As shown in \autoref{tab:gnn_content}, existing GNN education resources focus on explaining three main concepts: the data used in a GNN model, the computation process within a GNN layer, and the GNN model architectures. 
For data in GNNs, all resources explain graph structure, with 15/17 also covering high-dimensional node features.
All 17 resources cover the typical inputs and outputs of a GNN layer, with most delving into key computational processes, including neighboring node aggregation (14/17), neighboring node weights (13/17), and neighboring node sampling (9/17).
For overall GNN architecture, these resources cover GNN variants such as GAT~\cite{velivckovic2018gat}, and GraphSAGE~\cite{hamilton2017graphsage}, tasks such as link prediction and node classification, and the use of non-GNN layers such as MLP and global pooling.
We validated this list of key concepts with our GNN experts and incorporated all of them into the design and development of \name.

The reviewed materials employ a variety of formats to explain these key concepts of GNNs, including mathematical formulas, abstract diagram, python code, and data visualizations.
Mathematical formulas, such as \harry{the formula for message passing operation in graph convolution layer} $x_i = \sigma( W\sum_{j\in N(i) \cup i} \frac{e_{i, j}}{\sqrt{d_i d_j}} x_j + b )$, provide a precise and concise representation of the computations of GNNs. 
These mathematical formulas are often paired with abstract diagrams to enhance accessibility and intuitive understanding.
Given Python's prominence in GNN model development, code blocks are widely (10/17) used to demonstrate a GNN model is implemented in Python by constructing various layers and their connections.
It is worth noting that visualizations showcasing the internal mechanisms of GNNs using real-world data are relatively rare (4/17).

\subsection{Design Goals} \name\ aims to complement existing GNN educational resources by offering comprehensive and intuitive interactive visualizations of GNNs' inner workings.
\qw{The primary target users are novice AI practitioners with prior knowledge of mathematics and statistics (\eg, computer science students). 
At the same time, similar to other educational visualization tools~\cite{wang2020cnnexp, guo2013online}, \name\ can also support instructors who teach GNNs.}
The design goals were shaped by the key challenges in teaching GNNs reported by four GNN experts, and limitations identified in existing educational materials.

\begin{enumerate}[label=\textbf{G.\arabic*}, leftmargin=*, labelwidth=!, labelindent=0pt]
 
    \item 
    \label{goal:integrate_diverse_computation}
    \textbf{Integrate Diverse Computations}: A comprehensive understanding of GNNs encompasses a wide range of computations, including the graph data structure, the aggregation of node neighbors,  as well as non-GNN layers such as MLP and global pooling (\autoref{tab:gnn_content}). 
    Although these computations have been covered by the surveyed tutorials, they are often explained in separate sections.
    Therefore, it can be challenging for learners to see how these various computations interact with each other.
    

    \item \label{goal:demystify_math}
    \textbf{Demystify mathematical formulas}:
 
    As with most AI models, GNNs involve complex mathematical functions that define their structure and operations. 
    GNN \harry{tutorials} have a heavy reliance on mathematical formulas to describe their computations, as observed in 15 out of 17 reviewed resources.
    \qw{While precise, concise, and maybe preferred by users with strong math knowledge, these formulas can be overwhelmingfor beginners to effectively grasp.
    Even though most of the surveyed resources attempt to aid comprehension with diagrams and annotations, they primarily focus on defining individual terms (\eg, $W, b$) rather than linking these mathematical formulas with the actual data transformations occurring within a GNN. }
    According to the discussion with the GNN experts , such connections are crucial for learning GNNs.
    As E2 noted, it is about \textit{``translating the equations into a mental model of the data transformations''}.
    

    \item
    \label{goal:real_data}
    \textbf{Connect abstract concept with real data}: 
    \qw{Direct exploration of real-world examples is crucial for learning ML models, both to understand why certain design choices (\eg, pooling, sampling) are needed, and to ground abstract mathematics in concrete tasks (\eg, classify a molecular structure).}
    However, most existing tutorials (14/17) rely on simplified diagrams (\eg, an illustration of a 4-dimensional vector) for demonstrating the computation processes. 
    While some tutorials (10/17) use Python code to access real data, only 4 visualize real input–output data, and none visualize the internal workings of GNNs.
    \harry{We have included those four visualization in the supplementary material.}
    \qw{GNN experts emphasized that, beyond toy datasets for introducing new concepts, real-data visualizations are essential for bridging theory and practice. 
    This principle is also reflected in many widely used AI education tools \cite{wang2020cnnexp, cho2025transformer, LLMVIS}, which all use complex real world data and show the complete computation upon them.}
    
    
    %

    \begin{figure*}
    \centering
    \includegraphics[width=0.95\linewidth]{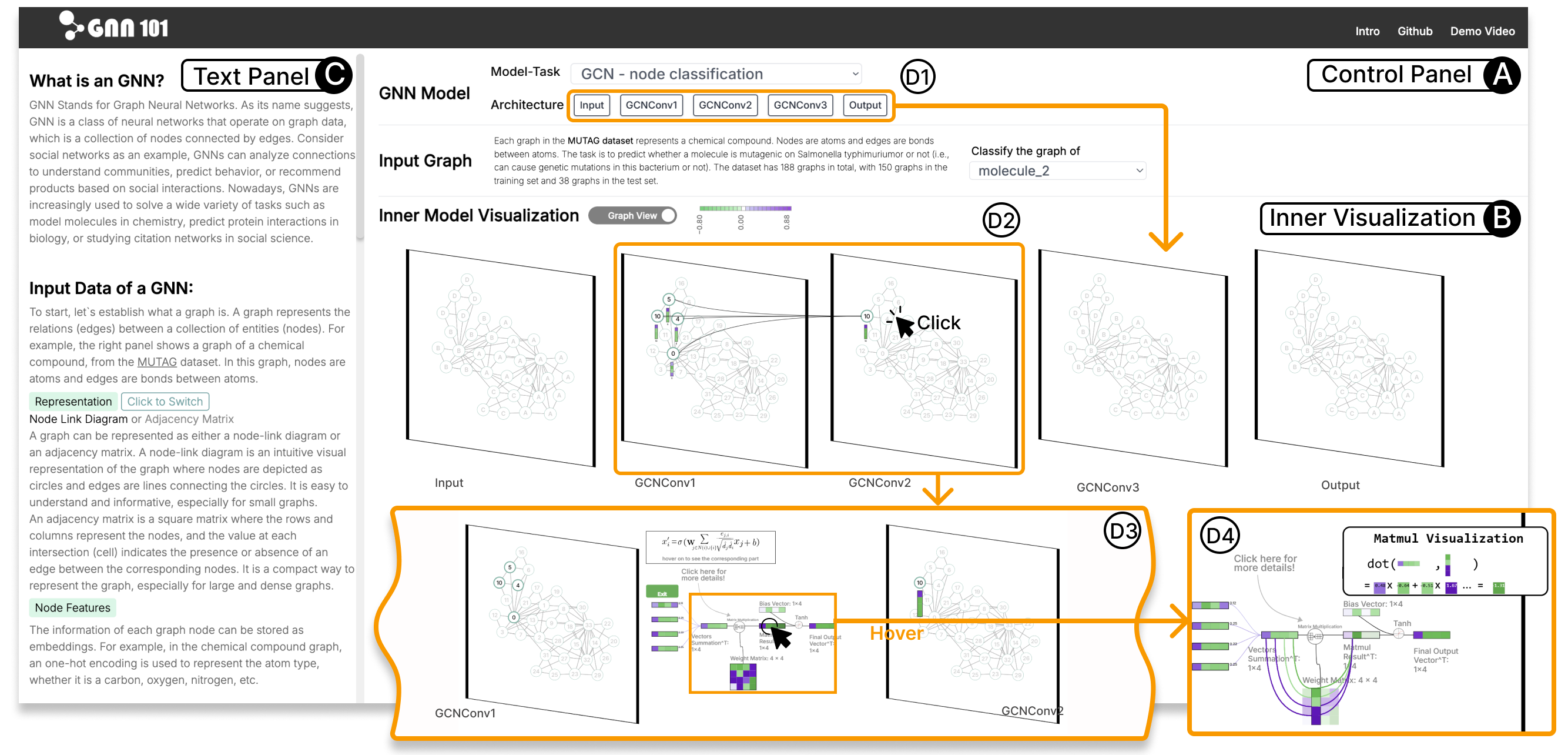}
    \caption{\textbf{The interface of \name{} in graph view}: (A) a control panel for selecting GNN models, tasks, and dataset; (B) an inner model visualization that displays the GNN model's inner workings; (C) a text panel that guides users in interacting with the visualization (C). The visualization provides hierarchical levels of detail, \harry{including a model overview(D1), layer-by-layer outputs (D2), click a layer to reveal the message passing (D3), and hover elements to display the associated mathematical operations and data types (D4).}}
    \label{fig:interface}
\end{figure*}

    \item 
    \label{goal:engagement}
    \textbf{Effective communication and active engagement}:
    In spite of the recent popularity of using visualization to explain ML to beginners~\cite{wang2020cnnexp, olah2017feature}, the educational benefits of visualizations are not always guaranteed. 
    \qw{Prior studies~\cite{hundhausen2002meta, gurka1996testing} showed that the success of visualizations in education depends on meaningful engagement. Passive viewing shows little advantage over conventional materials, whereas interacting with visualizations yield more effective learning outcomes.
    }
    \qw{Therefore, the successful educational use of visualizations requires active student engagement. }

\end{enumerate}

\section{Visualization Interface}


The interface of \name\ (\autoref{fig:interface}) consists of a control panel for selecting different datasets and GNN models (A), 
an inner model visualization that displays GNN's inner workings (B), 
and a text panel with an on-boarding tour (C).
This section introduces the main features of \name\ and explain how they achieve the design goals outlined in \autoref{sec:design}.

\subsection{Hierarchical Levels of Details}
\label{subsec: levels-of-details}

\name\ integrates various hierarchical levels of detail, ranging from the overall model architecture to the intricate data transformations within a specific layer (\ref{goal:integrate_diverse_computation}).
By revealing computation processes hierarchically, \name{} enables effective visualization of large-scale data involved in a GNN without overwhelming users (\ref{goal:real_data}).

First, users can observe the architecture overview next to the GNN model and task selectors, as shown in \autoref{fig:interface}.D1.
Different GNN tasks (\eg, a node or graph classification) and GNN variants (\eg, graph convolution or graph attention) will lead to different model architectures.

Second, users can view a more detailed visualization that shows the outputs of each layer. 
Clicking on a layer in the architecture overview highlights the corresponding layer in the detailed visualization.
As shown in \autoref{fig:interface}.D2, 
hovering over a node in a given layer highlights the relevant nodes in the previous layer that contribute to its feature computation. 
This interaction illustrates how node features are progressively learned from their neighbors, layer by layer.

Third, when users click on a specific layer output, \name\ visualizes the detailed computation within that layer in the style of a horizontal flow chart, as shown in \autoref{fig:interface}.D3. 
Upon selection, the chosen layer expands, while non-selected layers shift to the side and fade in opacity. This animated transition and the focus+context design helps users focus on the selected layer while still preserving the overall context of the GNN model.
In the flow chart, the inputs, outputs, and learnable parameters of this layer are visualized as heatmaps (\autoref{fig:matrix-view}.B1), where the shape represents the vector's dimensions, and the cell colors indicate their respective values.
Internal results are also displayed as heatmaps to break down a complex computation process into multiple steps for easy interpretation.
Curves connecting these heatmaps illustrate the computation process.
Icons on the curves indicate specific types of computations (\autoref{fig:matrix-view}.B2), while curves without icons represent the addition of factors.
Thickness and color of these curves encode the corresponding multiplication factors.

Lastly, to gain a deeper understanding of each step in the computational process within a layer, users can hover over a cell in the heatmaps to reveal the specific calculations that determine the selected cell's value, as shown in \harry{\autoref{fig:interface}.D3-D4}. 
In the pop-up windows displaying these computations, we apply the same color encoding of the heatmap to the background of each number, reducing cognitive load when switching between levels of details. 



\subsection{Complementary Views}
\label{subsec: two-views}
\begin{figure}
    \centering
    \includegraphics[width=\linewidth]{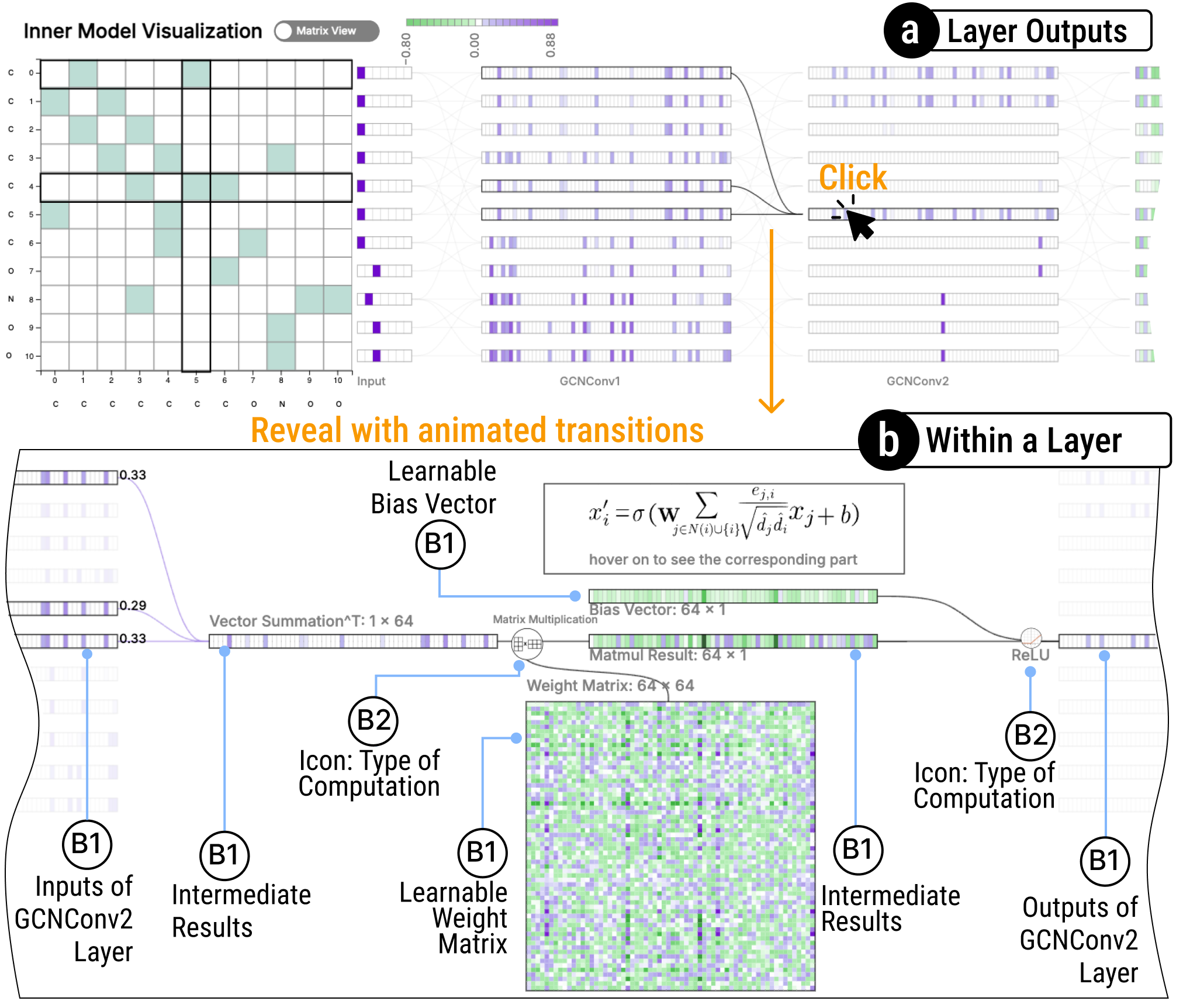}
    \caption{\textbf{Matrix View}: The matrix view complements the node-link view (\autoref{fig:interface}) and provides similar click-to-expand interactions (a-b). The computation process inside a layer is visualized as a horizontal flowchart, where heatmaps represent vectors and matrices (B1), and the connecting curves illustrate the computation process (B2). }
    \label{fig:matrix-view}
\end{figure}

In the second hierarchical level of details (\ie, layer inputs and outputs), \name\ provides both a node-link view (\autoref{fig:interface}) and a matrix view (\autoref{fig:matrix-view}) for visualizing GNN's inner workings. 
\qw{Large-scale user studies have consistently shown that node-link and adjacency matrix representations offer complementary strengths for different graph analysis contexts~\cite{ren2019understanding, ghoniem2004comparison, okoe2015graphunit, ghoniem2005readability}. Node-link diagrams offer intuitive representation of network topology, and provide superior task accuracy and completion times for small to medium-sized graphs. 
In contrast, matrix representations excel when analyzing larger or denser graphs, avoiding the visual clutter that can compromise node-link readability, though they often present a steeper learning curve. 
Therefore, \name{} provides both views, enabling them to complement one another while easing the learning curve of adjacency matrix.}

In the node-link view, the input graph and the intermediate layer outputs are visualized as node-link diagrams, as shown in \autoref{fig:interface}.B.
\harry{This view offers an intuitive representation of the graph structure and illustrates the key concepts of a GNN model, \ie, updating node features layer by layer, which helps users connect abstract concepts with real datasets (\ref{goal:real_data}).}
Users can hover over a node to examine its feature at the corresponding layer. 
Each feature is visualized as a heatmap, with the color of each rectangle indicating the dimension's value.
At a certain layer, a node's feature is computed based on its feature and those of its neighboring nodes from the previous layer.  
Such connections between the selected node and relevant nodes from previous layers are also highlighted when hovering.

In the matrix view, the input graph and the intermediate layer outputs are visualized as matrices, reflecting the exact format in which they are processed within a GNN model.
The graph structure is visualized as an adjacency matrix, where each row and column represent a node, and each cell indicates whether the corresponding nodes are connected. 
Node features are visualized similarly to those in the node-link diagram (\ie, heatmaps), but are arranged horizontally to align with the corresponding rows in the adjacency matrix. 
Unlike the node-link view, where node features are displayed only upon hovering, the matrix view presents all node features for each layer simultaneously to offer a comprehensive overview (\ref{goal:integrate_diverse_computation}).
Users can hover on a node feature to highlight its connections with the relevant node features from the previous layer, mirroring the interaction in the node-link view.


The two views complement each other: 
While the node-link view offers node-link diagrams for an intuitive representation~\cite{ren2019understanding}, the matrix view provides a comprehensive overview of features and shows data in the exact format used in GNN models.
Meanwhile, \name\ facilitates smooth transitions between two views via consistent interactions (\ie, hover over and click) and visual encoding (\ie, color encoding).


\begin{figure}
    \centering
    \includegraphics[width=0.9\linewidth]{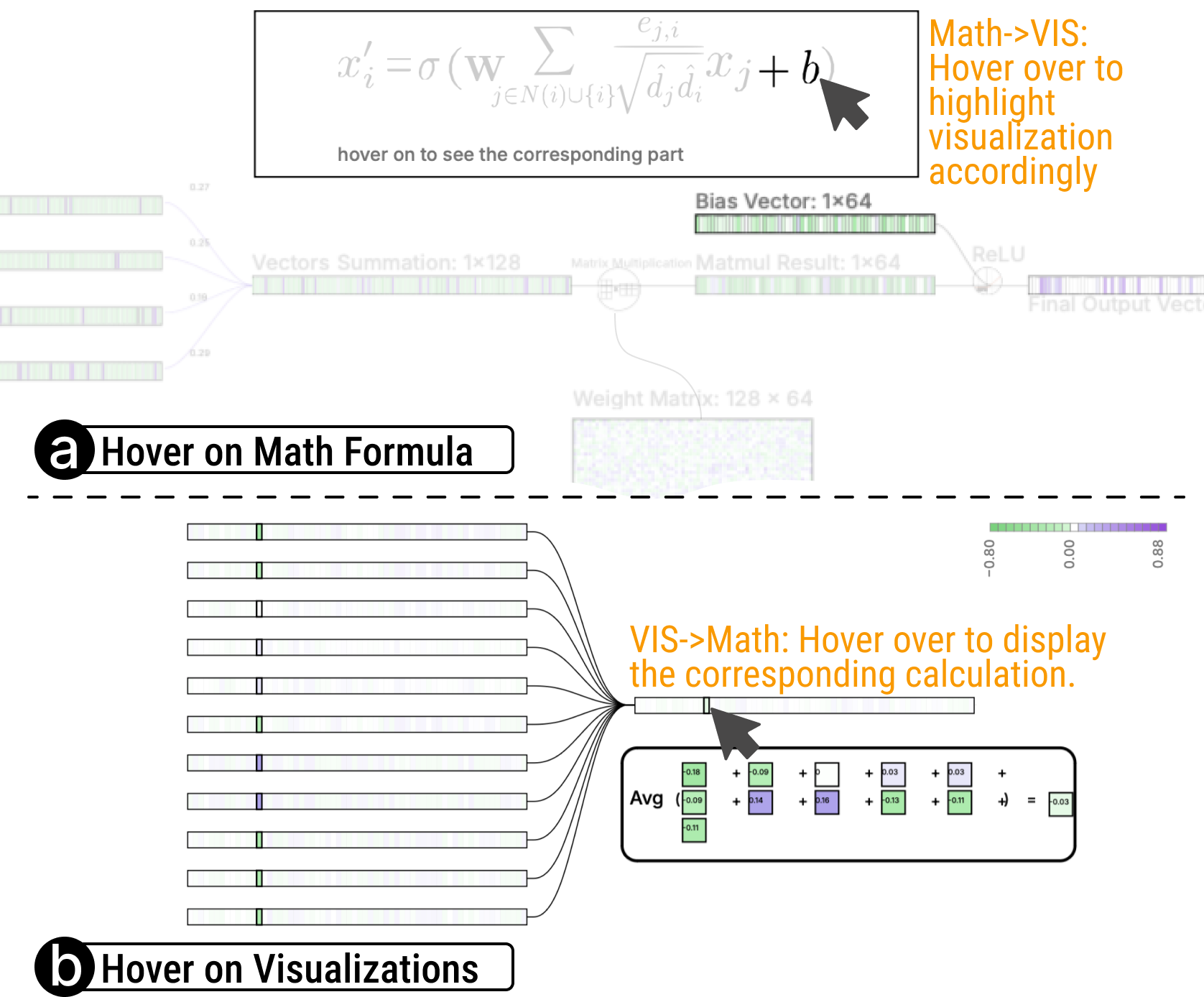}
    \caption{\textbf{Bidirectional Math-Visualization Linking:} Users can hover over parts of the mathematical formulas to highlight the corresponding visualizations (a), or hover over visualizations to reveal the computation process for obtaining the exact value (b).}
    \label{fig:math-linking}
\end{figure}

\subsection{Interactive Math-Visualization Linking}
\label{subsec: math-vis linking}


\name\ offers interactive bidirectional linkage between mathematical formulas and their corresponding GNN visualizations, aiding in the interpretation of complex mathematical concepts~(\ref{goal:demystify_math}).
When users click to expand a layer, the corresponding mathematical formula appears above the visualization. Users can hover over mathematical notations in the formula to highlight the corresponding parts in the visualizations. For example, as shown in \autoref{fig:math-linking}.a, hovering over the $b$ symbol in the formula highlights the bias vector in the flow chart.

Conversely, hovering over visualizations, such as cells in heatmaps, reveals detailed mathematical calculations of exact values, as shown in \autoref{fig:math-linking}.b.
The calculation using actual values provides concrete examples for a better understanding of abstract mathematical formulas.

\chongwei{\name\ provides such visual linking for most important components in the formulas, such as weight multiplication, activation function, aggregation, attention mechanism, and dot products, omitting the detailed yet minor components to avoid confusing or overwhelming users. }
Since this hover interaction only explains the calculation of a single dimension of a high-dimensional node feature, \name\ also provides an animation that demonstrates the step-by-step calculation for all dimensions. 

\subsection{Animated Transitions and Text Guidance}
\label{subsec: engagement}
Animated transitions are incorporated to guide user attention and enhance engagement (\ref{goal:engagement}) by ensuring smooth navigation between hierarchical levels of detail and gradually introducing changes to prevent cognitive overload.
Following the \textit{congruence} and \textit{apprehension} principles of animated data transitions ~\cite{heer2007animated}, we group similar transitions and stage complex ones for clarity.
For example, when the flowchart appears (\autoref{fig:matrix-view}.a to \autoref{fig:matrix-view}.b), the other layers first fade out while the input and output nodes maintain their opacity. 
The flowchart is then revealed step by step to connect the input and output nodes.
The revealing order aligns with the computation order: neighbor aggregation, weight multiplication, bias addition, and activation function processing. 
This structured approach helps users follow intuitively how data is transformed at each step.

In addition, text hints (\eg, \textit{click here for more details}, \textit{hover on to reveal the computation}) are added to guide users in navigating the interface, explaining which interactions are available.
We also use flash animations to direct users' attention to important areas, such as the \textit{Click to Predict} button on the landing page.

\subsection{Data, Task, and Model Exploration}
\label{subsec: exploration}
In the control panel,
users can easily explore different GNN models for various tasks and input graphs~(\ref{goal:real_data}).
As it is impractical to include all possible options, we selected representative examples based on the 17 surveyed tutorials.
These design choices were also validated or modified according to the discussion with the GNN experts.

\noindent
\underline{Data:} \name\ provides \qw{one synthetic toy dataset and three real-world graph datasets: chemical compound graphs~\cite{mutag}, a social network of a Karate club~\cite{zachary1977information}, and a social network of Twitch players~\cite{rozemberczki2021twitch}.}
\qw{The synthetic toy dataset allows users to begin with simple models and data (\eg, embeddings with lower dimensions and fewer GNN layers), enabling them to first grasp the foundations of GNNs through straightforward examples. 
%
The real-world datasets cover diverse applications, with graph sizes ranging from small graphs (\eg, five nodes) that are commonly used for graph classifications to large graphs (\eg, hundreds of thousands of nodes) that are commonly used for link/node predictions.
The synthetic toy dataset also enables direct comparison across GNNs with identical weights, parameter configurations, and feature dimensions, complementing the real-world datasets that naturally exhibit architectural GNN variations due to the fundamental design requirements of different GNN tasks. 
}

\noindent
\underline{GNN Variants}: \name\ covers three most widely-used GNN variants, Graph Attention Networks (GAT)~\cite{velivckovic2018gat}, Graph Convolution Networks (GCN)~\cite{kipf2022GCN}, and GraphSage~\cite{hamilton2017graphsage}.
\qw{\name\ provides a list of pre-trained models among these GNN variants to cover the diversity of common architectural design choices, including different aggregation functions, activation functions, weight configurations, and sampling strategies. 
By comparing authentic models with different design choices, users can gain genuine insights into how architectural decisions impact model behavior and computational flow, while avoiding the technical complexities and potential confusion that unconstrained parameter editing would introduce.
}

\begin{figure}
    \centering
    \includegraphics[width=1\linewidth]{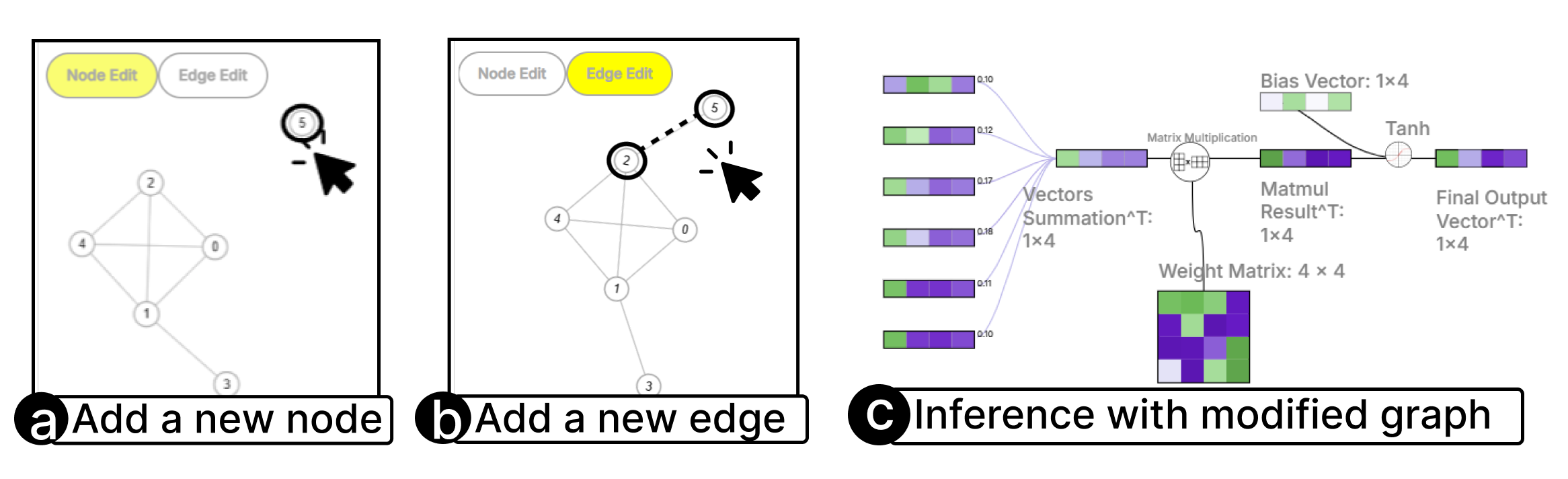}
    \caption{\harry{\textbf{Manipulating Input Data using Graph Editor:} Users can manipulate the input data using a graph editor to choose between node operation (b) or edge operation (c) in the editor menu (a), after the operations, users can click the "Click to Predict" button to run inference with the new modified data (d). }}
    \label{fig:graph-editor}
\end{figure}

\noindent
\harry{\underline{Direct Manipulations on Input Data}: 
Direct manipulation of input data enhances learning by enabling users to immediately observe how changes affect model behavior. As shown in \autoref{fig:graph-editor}, \name\ integrates a graph editor that allows users to directly modify input graph structures. These modifications will be reflected with the main visualizer, and users can trigger model inference with the updated graph by clicking the ``Click to Predict'' button. 
This interactive approach encourages exploratory learning, where users can experiment with different graph configurations and immediately observe how variations in input structure influence the GNN's internal computations and final predictions. 
} 


\subsection{Implementation}

\name\ includes pre-trained GNN models and interactive visualizations of these models.
The \textbf{GNN models} are pre-trained in Python using Pytorch Geometric~\cite{pyg}, 
covering various graph datasets, tasks, and GNN variants.
These pre-trained GNN models are then exported into ONNX format.
The \textbf{visualization} is implemented in TypeScript using React and D3.js~\cite{d3}. 
The pre-trained GNN models are loaded in the user's web browser using the ONNX Web Runtime~\cite{onnxruntime}, which enables \harry{real-time graph editing and inference computation}.
\harry{The current version of GNN101 supports a wide range of customizations, including adaptation to new GNN models with varying weight dimensions and neural network architectures. The source code, documentation for running and extending the system, and a web-based demo are available at \url{https://github.com/Visual-Intelligence-UMN/GNN-101}.
 }

\section{Use Cases}
We deployed \name\ in three GNN-related courses at three different universities.
Similar to other educational visualizations~\cite{wang2019genealogy, guo2013online}, we have observed diverse use cases, including its use by instructors as a teaching aid in their lectures, 
by teaching assistants (TAs) to assist students during office hours, 
and by students for independent review of course materials after class.
In this section, we show how \name\ can help a computer science student, Alice, learn various concepts in GNNs based on the reported usage. 
Even though we present Alice as the primary user, the same cases apply to instructors and TAs in teaching GNNs.

\subsection{Understanding Message Passing}
Alice wants to gain a deeper understanding of message passing, the fundamental mechanism underpinning most GNNs~\cite{velivckovic2018gat, kipf2022GCN, sanchez-lengeling2021gentle}. 
However, the lecture slides and online GNN educational materials primarily explain this concept using diagrams and mathematical formulas, as shown in \autoref{fig:teaser}.a, which Alice finds less intuitive and effective. 
To enhance her learning, Alice turns to \name\ for assistance.

\harry{Alice first starts with the matrix view, which visualizes the input data and output data as a matrix and vectors (heatmap), helping Alice then understands the basic data formats. }
\harry{However, Alice gets lost about the computation between layers, so she changes to the node-link view. }
When Alice hovers over a node in the GCNConv2 layer output, \name\ highlights the selected node, along with its connections to itself and neighboring nodes in the previous layer’s (GCNConv1) output. 
This interaction helps Alice intuitively grasps the high-level concept of message passing: updating a node’s features by aggregating information from both its neighbors and itself (\autoref{fig:interface}.a).
Curious about the detailed computations of message passing in GCN, Alice clicks on the node to expand the layer and reveal the calculation process as a horizontal flowchart (\autoref{fig:interface}.b). 
A mathematical formula of the calculation also appears to facilitate the understanding. 
When Alice hovers over any symbol in the equation, the matching element in the flowchart is highlighted.
This helps Alice better understand understand both the expected data formats and how each component contributes to the computation.

Lastly, to refresh her memory on matrix multiplication and the ReLU activation function, Alice hovers over the results of these calculations, triggering tooltips that explain the process and demonstrate how specific values are computed (\autoref{fig:interface}.D4).
Alice also observes the representation power of GNNs by directly modifying the input data through the graph editor (Figure \ref{fig:graph-editor}) and seeing how GNNs update the node features layer by layer to generate accurate predictions.



\begin{figure}
    \centering
    \includegraphics[width=\linewidth]{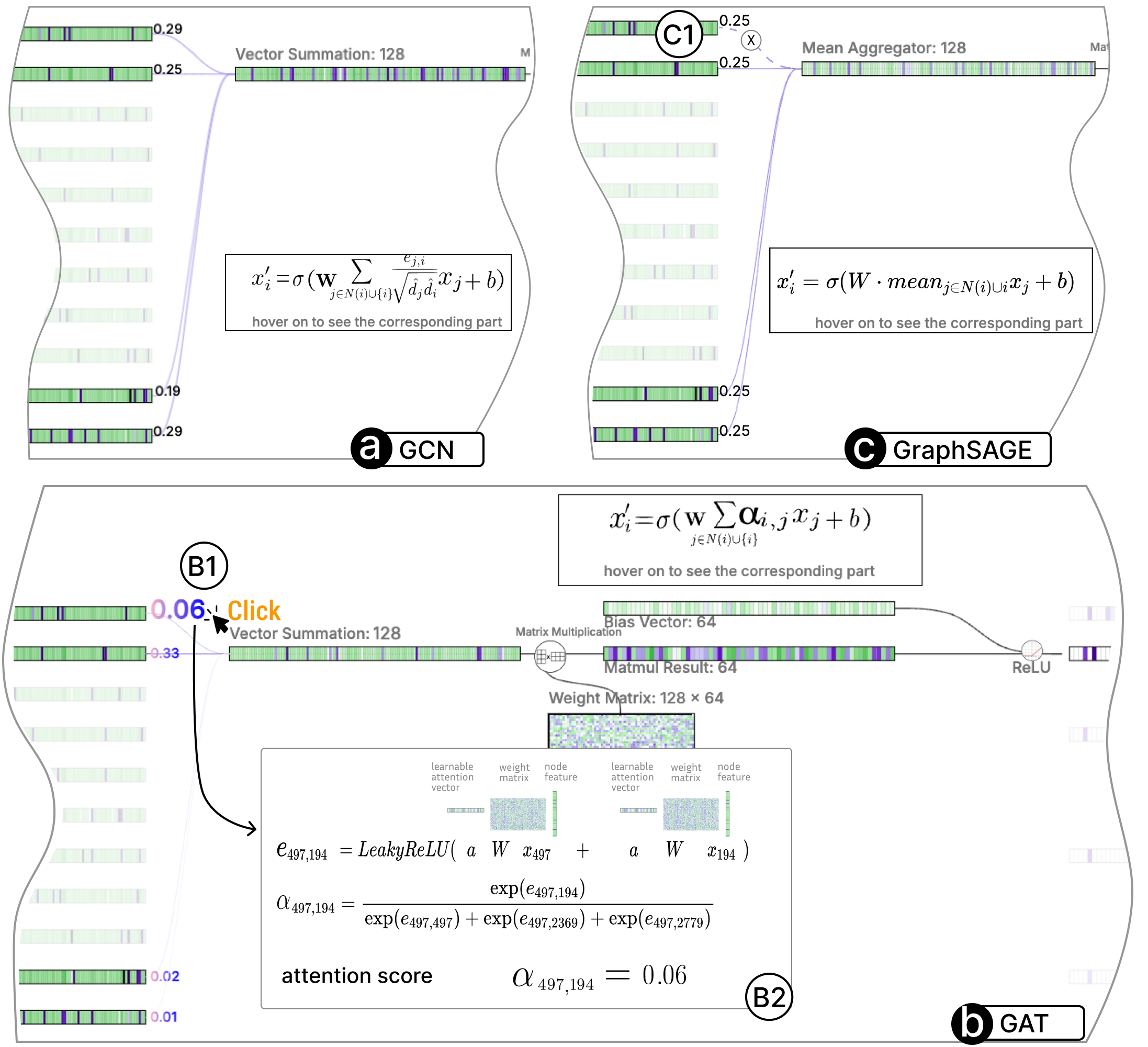}
    \caption{\textbf{Comparing GNN variants}: \name\ supports the comparisons between GCN (a), GAT (b), and GraphSAGE (c) by visualizing how each model aggregates information from neighboring nodes. Curved edges indicate that parts of the visualization have been omitted from the figure due to space constraints.}
    \label{fig:gnn-variants}
\end{figure}

\subsection{Comparing GNN Variants}
After understanding message passing and its implementation in GCN, 
Alice turns to \name to examine other two GNN variants: GAT and GraphSAGE.

Alice starts by selecting GAT in the control panel and clicking on a node in a GATconv layer to view the internal computation inside this layer.
The computations are visualized using a horizontal flowchart similar to the one for GCN but with differences in how the input features are connected and aggregated. Unlike GCN, which uses the node degree $\frac{1}{\sqrt{d_i d_j}}$ as an aggregation factor, GAT computes an attention score $\alpha_{i,j}$, which is visualized with gradient colors for distinction and interactivity. 
As shown in \autoref{fig:gnn-variants}.B2, Alice clicks on an attention score, revealing a step-by-step breakdown of the computation through three mathematical formulas. 

Next, Alice selects GraphSAGE in the control panel and clicks on a node in a SAGEconv layer to view the internal computation.
Sampling icons (\autoref{fig:gnn-variants}.C1) appear on the curves connecting and aggregating input features, highlighting the neighbor sampling process, a key innovation of GraphSAGE. 
Clicking on these icons triggers tooltips that provide textual explanations of the sampling method used in GraphSAGE.
Alice concludes that the primary differences among the three GNN variants lie in how they select and aggregate information from neighbors, which are visually represented by the curves connecting input features to intermediate summation results (\autoref{fig:gnn-variants}.a, B1, and c).


\subsection{Learning Different GNN Tasks}
After understanding message passing and its implementation across GNN variants, Alice is eager to explore how GNNs solve various tasks in real data. 
In \name, Alice uses the control panel to switch between three common GNN tasks: graph classification, link prediction, and node classification. 
She notices that the first several layers of GNNs for these tasks are very similar.
Each layer updates the graph node features by learning from its neighbors.


\begin{figure}
    \centering
    \includegraphics[width=\linewidth]{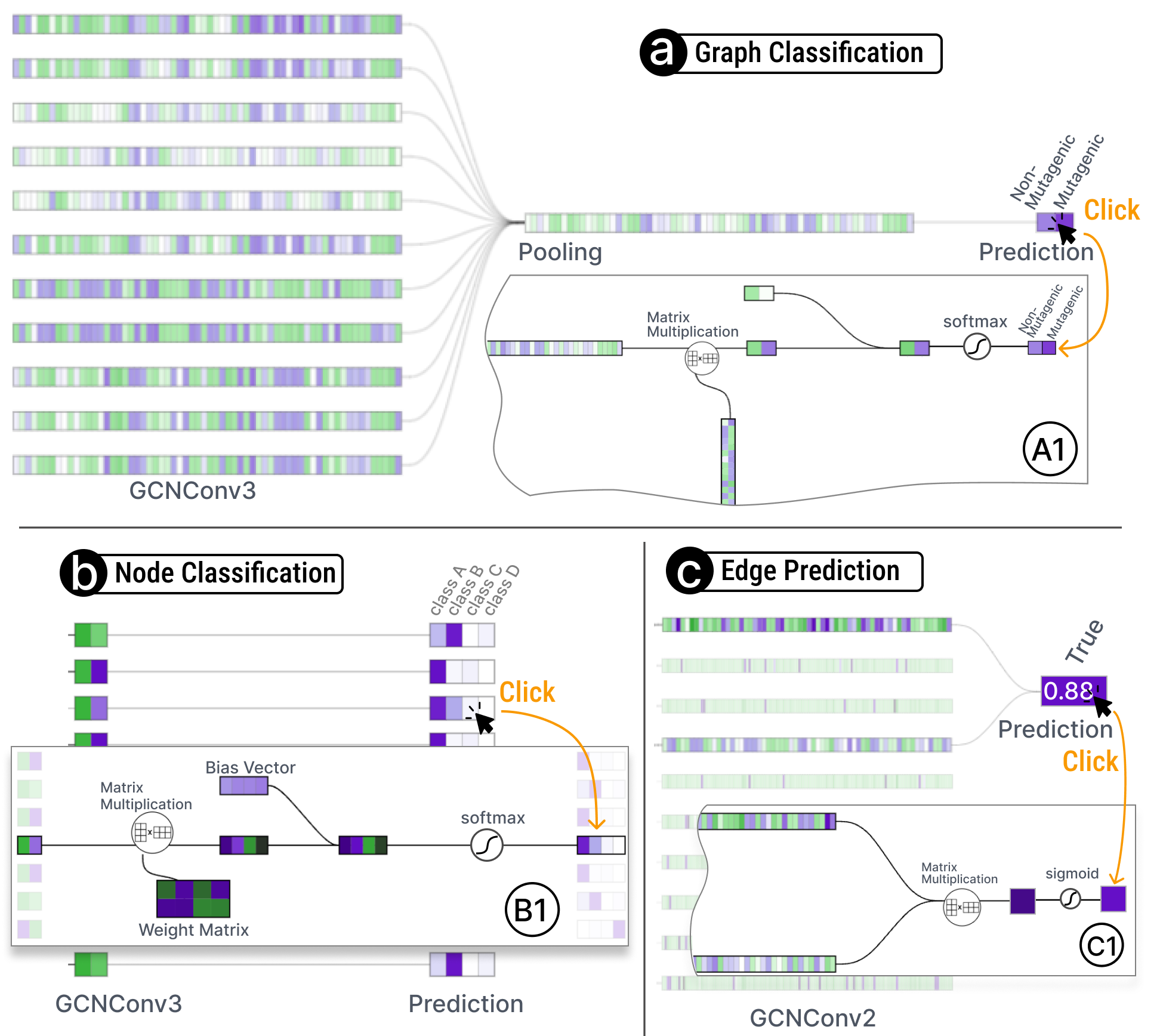}
    \caption{\textbf{Learning different GNN tasks:} The final one or two layers handle task-specific processing for graph classification (a), node classification (b), and edge prediction (c). These additional layers also support the same click-to-expand interactions as the GNN layers. Curved edges indicate that parts of the visualization have been omitted from the figure due to space constraints.}
    \label{fig:task}
\end{figure}

Alice observes the main differences occur in the final one or two layers for task-specific processing (\autoref{fig:task}). 
In graph classification (a), a global pooling layer aggregates all node features of the graph and outputs a single feature, which is used to generate the graph classification via an MLP layer. 
Alice hovers over this layer to see the detailed computation: an element-wise average across dimensions. 
In node classification (\autoref{fig:task}.b), an MLP layer is applied to each node feature to generate predictions. 
In edge prediction (\autoref{fig:task}.c), the features of the two nodes involved in the predicted edge are fed into a prediction layer that outputs a prediction score. Alice clicks on the prediction layer and sees a dot product of the two node features used (\autoref{fig:task}.C1).

\subsection{Iterative Refinement Through Classroom Deployment}
\label{subsec:feedback}
Throughout the deployment, we gathered feedback through questionnaires shared on the course discussion forum and informal interviews with TAs, instructors, and students who volunteered to provide additional input. 
This feedback directly informed several iterations.

A comment from a TA on the initial version, \textit{``we use a lot of mathematical formulas when teaching GNNs''}, motivated one of \name’s key features: the math-visualization linking. This feature enables users to hover over formulas and see the corresponding components highlighted in the visual pipeline, effectively bridging symbolic and visual representations. In later feedback, this feature was consistently praised.

One common piece of feedback we received from the initial version was that users were unsure where to begin and some students found the visual representation of the adjacency matrix confusing. This prompted us to introduce an onboarding tutorial that guides users through the interface and highlights key entry points. We also added a flashing animation to draw users' attention to the ``Click to Predict'' button, helping them initiate their first interaction with the tool. On the landing page, we integrated a coordinated node-link diagram and adjacency matrix view to familiarize users with both representations. Additionally, we incorporated text annotations indicating available interactions (\eg, ``click here'' or ``hover here'') to improve user experience.

Another area of feedback focused on the level of visual detail and the abruptness of changes when expanding a layer. One instructor commented, \textit{``It’s expected, since there’s a lot going on within one GNN layer, but the visualization can be overwhelming at first.''} 
In response, we introduced step-by-step animations that gradually reveal the computations within each layer, helping users process the information at a more comfortable pace. 
Additionally, we implemented animated transitions to smoothly bridge different levels of granularity, making it easier for users to maintain context as they navigate between high-level overviews and low-level operations.

Finally, \qw{based on feedback requesting simpler examples, we modified the dataset selection to include both small graphs (\eg, with 5 nodes) for concept illustration and larger graphs (\eg, with thousands of nodes) for demonstrating scalability and real-world relevance.}

\section{Evaluation}
\qw{In this section, we evaluate \name{} through three approaches: a quantitative comparison study demonstrating educational effectiveness, a qualitative observational study examining actual user interactions and experience with \name{}, and an analysis of user engagement in real-world settings.}

\subsection{Quantitative Study on Educational Benefits}

\qw{This study investigates i) whether \name{} improves users’ knowledge of GNNs, and ii) whether such improvement exceeds the existing baseline method. }

\subsubsection{Participants}
\qw{
We recruited 18 student participants (16 male, 3 females, age: 23.5±2.50) from a large university via email list. 
These participants have some knowledge of basic machine learning concepts (average self rated 3.77 on a 7-point Likert scale) but little no experience with GNNs. 
None of the participants in this group has prior knowledge about \name{}.
Each participant received a \$10 Amazon gift card.
}

\subsubsection{Procedure}
\qw{
We conducted a between-subject study comparing \name{} with an established interactive GNN article~\cite{daigavane2021convolution} as baseline. 
We used a between-subject design to prevent learning effects from multiple exposures. Participants were randomly assigned to conditions, with no significant differences in self-reported ML or GNN knowledge between groups.
We measured learning outcomes using pre- and post-study assessments with a 15-questions multiple-choice quiz that covers fundamental GNN concepts. Questions were developed through content analysis of existing GNN tutorials (see \autoref{tab:gnn_content}). 
In the baseline condition, participants read the interactive Distill article. In the experimental condition, participants used \name{} with access to a provided tutorial explaining the system's functionality. Sessions lasted up to 40 minutes or until participants indicated completion. We measured learning gains as the difference between pre- and post-quiz scores.
The provided tutorial for \name{} and the complete quiz are available in supplementary materials.
}

\begin{figure}
    \centering
    \includegraphics[width=0.95\linewidth]{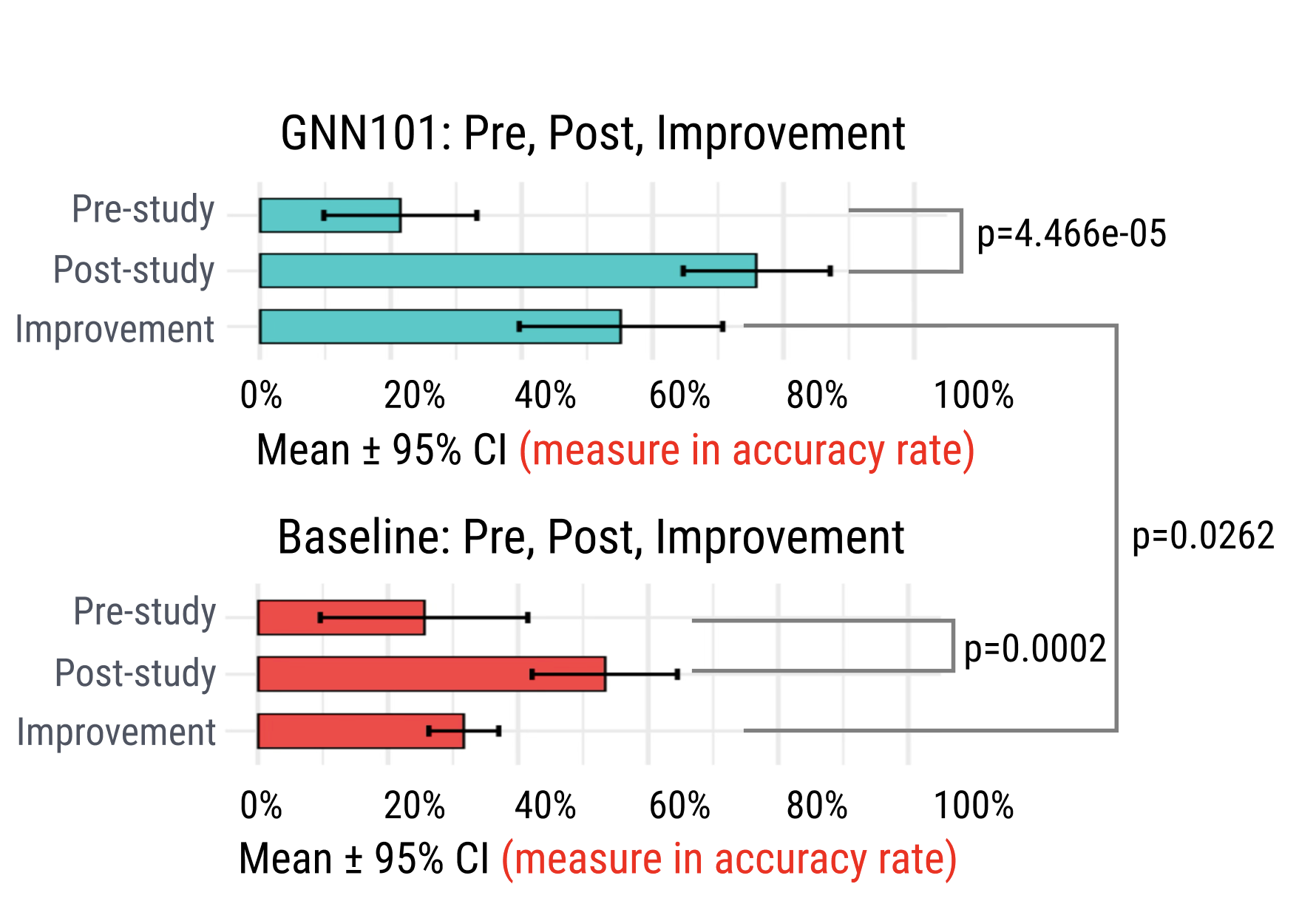}
    \caption{\textbf{Results of the Comparative Study}: The results of pre-quiz and post-quiz and the improvement between the pre-quiz and post-quiz, where red bars indicate the mean of Distill users' score and blue bars indicate the mean of \name\ users' score. Black bars indicate the group's $95\%$ confidence interval. }
    \label{fig:quiz-accuracy}
\end{figure}

\subsubsection{Analysis, Sample Size, and Results}
\qw{
We assessed the educational efficacy using pre- and post-study quiz accuracy scores, as illustrated in \autoref{fig:quiz-accuracy}. 
Within-condition changes were analyzed using paired t-tests, while between-condition comparisons used analysis of covariance (ANCOVA), with post-study scores adjusted for pre-study performance.}
\qw{
Power analysis was conducted assuming a large effect size (Cohen's d $>$ 1.3), based on pilot study.  
This effect size indicated that 18 participants (nine per condition) would provide approximately 80\% power at $\alpha = .05$. 
}

\underline{Within-condition learning gains.}
\qw{In both conditions, the accuracy of the post-study quiz improved significantly compared with that of the pre-study quiz.
For \name, the average accuracy improved from $21.4\%$ (95\% CI: $11.7\%$) to $75.9\%$ (95\% CI: $11.2\%$) ($p<.001$, $t(8)=7.97$). 
For baseline, the average accuracy improved from $25.6\%$ (95\% CI: $15.9\%$) to $53.3\%$ (95\% CI: $11.2\%$) ($p<.01$, $t(8)=6.35$).
}

\underline{Between-condition comparison.}
\qw{ANCOVA revealed that \name{} produced significantly greater learning improvements compared to the baseline condition ($F(1, 15) = \text{12.97}$, $p < .05$, partial $\eta^2 = 0.46$). 
The average improvements using \name{} is $55.19\%$ (SD $20.21\%$), while the average accuracy improvements in baseline is $27.78\%$ (SD $13.12\%$).
This suggests that \name\ not only significantly improve students' knowledge about GNNs, but also shows superior performance improvements than an established baseline.
}

\subsection{Observational Study}
To obtain detailed insights into user experiences and interactions with \name{}, we conducted a qualitative study involving observations of students and teaching assistants using the system, complemented by follow-up interviews.



\subsubsection{Participants and Procedure}
\qw{The qualitative included two participant groups representing our target user populations: instructors who taught GNNs and begining AI practitioners with ML background but limited GNN knowledge.}
We recruited 14 participants in total (11 males, 3 females, average age 23.8) through snowball sampling. 
The first group consisted of seven teaching assistants (TA1–TA7) for GNN courses we collaborated in \autoref{subsec:feedback}. 
They all had experience explaining GNN concepts in interactive, conversational settings such as office hours, labs, or tutorials.
The second group comprised seven students with a background in machine learning but had not studied GNNs, labeled S1–S7.
%
None of the participants had prior knowledge of this project nor participated in the quantitative study.


\subsubsection{Procedure}
We conducted the study with participants one-on-one via Zoom. 
Each study began with a 5-minute introduction to \name\ and an overview of its key features. 
After the tutorial, participants freely explored \name\ in their own web browsers, following a think-aloud protocol. 
A manual outlining the main functions of \name\ was available for reference, and participants could ask questions at any time during the process. 
Each session concluded with a usability questionnaire, identical to the one used in CNNExplainer~\cite{wang2020cnnexp}, followed by a semi-structured interview. The sessions ranged in duration from 30 to 50 minutes, and each participant received a \$10 Amazon gift card.


\subsubsection{Usability and Usefulness}



\begin{figure}
    \centering
    \includegraphics[width=\linewidth]{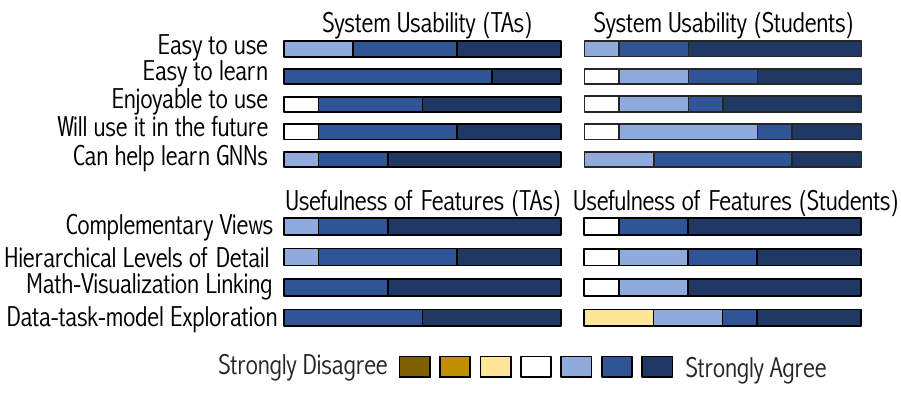}
    \caption{\textbf{Results of the post-study questionnaires}: Users rate the system usability and feature usefulness. Ratings were based on a 7-point Likert scale, where 1 indicates \textit{strongly disagree} and 7 indicates \textit{strongly agree}.}
    \label{fig:users-chart}
\end{figure}

\qw{We analyzed interview data following established qualitative analysis procedures, applying open coding and axial coding from grounded theory~\cite{strauss_corbin_1990} to identify emergent themes and patterns in user experiences.}

Overall, participants stated that \name\ addresses a strong demand for visualization when learning GNNs.
For example, TA6 stated \textit{``unlike images that look like grids, the unique structure of graph makes it hard to imagine how the data flows through the model... definitely a very useful visualization''}.
TA5 emphasized the importance of visual aids in GNN learning, stating, \textit{``understanding the core message passing steps of GNNs is pretty visual. I used to reference the animation in this distill.pub article (Sanchez-Lengeling~\etal~\cite{sanchez-lengeling2021gentle}) a lot when explaining things (to students)''}.
S2 reported \textit{``You guys are more interactive, more comprehensive (compared with \cite{sanchez-lengeling2021gentle})...I think it's pretty helpful, for sure''}.
TA7 commented \textit{``I think this is the best tool out there for learning GNNs''}.
Most of the participants (11/14) were able to conduct the free exploration without asking questions or referring to the manual.

This overall positive attitude is also reflected in the post-study questionnaires, as shown in \autoref{fig:users-chart}.
\qw{Both students (S1–S7) and TAs (T1–T7) gave overall positive feedback. More than five of the seven students (strongly) agreed on the usefulness of the proposed features, while all TAs (strongly) agreed. 
Within this generally positive trend, we observed that students and TAs diverged in their ratings on certain items.
For example, on \textit{``easy to use''}, students gave higher ratings comparable to those of TAs. 
In contrast, the \textit{``data–task–model''} exploration feature was rated as the most useful by TAs but the least useful by students. 
We suspect this discrepancy arises because many students were encountering GNN concepts for the first time through our visualizations, making some features more cognitively demanding. 
The variation in ratings across items appears consistent with the complexity of the underlying GNN concepts} \harry{and different students' preference to math formula and visualization.}
Notably, three participants mentioned that they would be continuing as TAs in the upcoming semester and asked permission to use \name\ for the students.


\subsubsection{Observations} We observed some consistent patterns in how participants interacted with \name.
 
Most participants (10/14) followed an overview-to-detail approach for exploration. They typically began by exploring the overall model structure, then selected specific nodes to observe the message passing, and finally hovered over individual components to examine the underlying mathematical operations. 
Such an interaction flow demonstrates the importance of the hierarchical level of details, 
as noted by TA4, \textit{``I like it ... if we present all of the information at once, it might be confusing to know what to focus on first''}.

We also observed a tendency to move from intuitive to concrete. 
\qw{All participants chose the node-link view to start their exploration, likely due to its default status and the intuitive representation of graph data.
As TA4 commented, \textit{``that's (message passing on graph view) super clear... when I worked as a TA, I found that is where many students had trouble understanding.''} 
Participants then switch to the matrix view to \textit{``see the outputs of all layers''} and \textit{``connect theoretical concept with practical GNN data flows''} (TA7).
}

\qw{Some participants initially struggled with visualizations or the math formulas, but found clarity through the linkage between them. 
For instance, S4 initially felt confused when viewing the message-passing visualization. 
However, S4 understood the visualization after interacting with the formula, which describes the message passing as the aggregation of neighboring node features. S4 remarked: \textit{``This is probably my favorite feature.''}
}


\subsubsection{Feedback} 
In the post-study interviews, participants, especially TAs, frequently compared \name\ with other GNN educational resources, 
and stated that \name\ presents a unique advantage, especially through the hierarchical levels of detail and the integration of mathematical formulas with visualization. S3 remarked: \textit{``make it even more useful than a good YouTube video.''} 

We also received valuable suggestions for improving \name{},
The most common suggestion we received was to incorporate additional features that support more advanced concepts in GNNs, \eg, inductive and transductive GNN algorithms, and the aggregation of both node and edge features. 
Other suggestions were mainly about minor changes such as making the control panel resizable and adding model training information.
These suggestions reflect participants' appreciation of the current version and their desire to extend \name\ to encompass broader concepts. 


\subsection{User Activities in the Wild}
\qw{As pointed out in the meta-study conducted by Hundhausen et al.~\cite{hundhausen2002meta}, active user engagement with visualizations has the greatest impact on their educational effectiveness.
We utilized Google Analytics\footnote{The website is available since August 10, 2024. Our analysis is based on data collected up to August 12, 2025.} to monitor the deployed \name{}.
We only considered \textbf{1,593 active users}, \ie, those who not only opened the website but also interacted with the visualizations. 
Since we had not publicly advertised \name, we initially expected that the users would primarily be instructors, TAs, and students from the three GNN-related courses we collaborated with (approximately 200 users out of more than 300 students, from three different states of the U.S.).
However, as the URL of \name\ was publicly accessible, we were surprised to find that over 1,200 users from 77 countries discovered and used it organically.
The fact that many users discovered and engaged with \name\ without any targeted promotion suggests its high adoption potential and strong engagement.
On average, each user performed \textbf{26.24 events} (\eg, click, scroll, hover), with an average engagement time of \textbf{5 minutes and 8 seconds} and a \textbf{33.83\%} return rate. }

\section{Discussion} 
This section discusses the broader design implications of our approach, highlighting how our visualization strategies can inform future tools and research in AI interpretability and education. 
It also outlines the current limitations and potential directions for future work.

\subsection{Design Leassons and Implications}

\noindent
\textbf{Animated Transitions in Educational AI Visualization.}
\qw{
An important design lesson from this study is the value of animated transitions for educational visualization, both for increasing engagement and for reducing information overload. 
We frequently received comments like \textit{``Aww, that is so cool!''} from both students and instructors when showing the animated transition features to them. 
Their immediate eagerness to try it themselves highlights the importance of animated transitions in capturing user interest and enhancing engagement.
Beyond engagement, animations that visually transport node features, progressively reveal matrix multiplications, and stage flowcharts step by step help learners maintain a coherent mental model across levels of detail.
Their effectiveness reflects Gestalt principles \cite{wertheimer1938gestalt} of continuity and common fate, which states that smooth, predictable movements reduce cognitive disruption.
It also aligns with the segmenting principle from multimedia learning~\cite{mayer2005cognitive}, which suggests that better learning is achieved when breaking down complex information into smaller, manageable chunks. 
While previous studies have shown the effectiveness of animated transitions in learning visualization~\cite{ruchikachorn2015morphing, heer2007animated} and animations have been adopted in several educational AI visualizations~\cite{wang2020cnnexp, cho2025transformer}, their role has been rarely articulated or systematically examined within the emerging VIS4ML community~\cite{yuan2021survey}.
This limited attention might be caused by the field’s current predominant emphasis on analytic rather than educational applications.
}

\noindent
\textbf{Math-Visualization Linkage.}
\qw{A second key implication of this study is the importance of tightly interweaving mathematical formulas into AI model visualizations. 
This seemingly straightforward linkage is often absent in current AI visualizations.
While a few examples of math-visualization linkage exist~\cite{daigavane2021convolution, LLMVIS}, 
these implementations are often limited.
They either only focus exclusively on detailed matrix calculations without broader conceptual context~\cite{LLMVIS}, or oversimplify complex high-dimensional mathematical operations by reducing them to scalar representations~\cite{daigavane2021convolution}.
Many existing tools position formulas and visuals as alternatives, but our findings show that learners benefit most when they can seamlessly navigate between the two. By enabling bidirectional linkage between visual components and symbolic notation, \name{} supports learning via a deeper connection between abstract equations and concrete operations. 
This observation aligns with Paivio's dual-coding theory~\cite{paivio1990dual} and Mayer’s cognitive theory of multimedia learning~\cite{mayer2005cognitive}, which posits that humans process information through distinct verbal and visual channels and that learning is enhanced when these channels are coherently integrated.
}


\noindent
\textbf{Consistent Visualization Languages for AI.}
\qw{
Finally, \name{} demonstrates the importance of consistent and familiar representation. We adopted heatmaps for all vectors (weights, biases, node features) and flowchart for the computation process, leveraging visual languages already familiar to most learners. 
Although we examined alternative, space-efficient visualizations, familiar representations were more effective by reducing the need to reinterpret new encodings and focus on conceptual understanding.
This benefit mirrors the success of interactive spreadsheet metaphors, such as Tom Yeh's AI by Hand in Excel~\cite{yeh2024fullstacktransformer}, which leverages the ubiquitous spreadsheet interface to enable learners to manipulate inputs, inspect intermediate computations, and observe outputs in a transparent, consistent manner. 
We need consistent visual languages that remain effective, scalable, and familiar. 
Establishing such shared visual languages will be essential for advancing educational AI visualizations and will require collective effort from communities.
}

\subsection{Limitations and Future Work}
Our current implementation of \name\ has shown promise in enhancing GNN education, but there are several areas for improvement.


\harry{First, while \name\ supports various mathematical formulas for GNNs, it provides limited adaptability and flexibility in linking mathematical expressions to their visual representations. For example, interacting with formula can be in both symbol-level (\eg, a single symbol such as $d_i$) and block-level (\eg, an entire block of symbol such as $\frac{1}{\sqrt{d_i d_j}}$), help users understand the interactions between different parameters. 
}

\harry{Second, while our evaluation demonstrates the immediate usability, effectiveness, and educational benefits of \name{}, it lacks longitudinal assessment to understand knowledge retention and skill transfer over time. 
In addition, although we demonstrate the user engagement in the wild, but it is still need a detailed interaction log analysis to understand the users' sense-making process and understand which components can contribute to successful learning at large scale. 
Future work will include longitudinal studies and enhanced interaction logs to better understand the usage of \name{} in real world. 
}

\qw{Lastly, as a long-term goal, we plan to evolve \name\ into a modular and collaborative framework that allows the community to contribute visualizations for diverse models, tasks, and datasets. 
Even though the current version of \name\ provides comprehensive coverage of the foundational concepts, user feedback revealed demand for advanced features (\eg, dynamic adjustment on formula parameters), reflecting the complexity of GNNs. 
This modular and collaborative framework will ensure the compatibility of \name\ with other explorable-based learning. For example, an integration through ipywidgets can integrates \name\ into Jupyer notebook style tutorials which can help users understand GNNs in a different format and closer to real-world work settings.  
}

\section{Conclusion}
This study proposes \name, an interactive visualization tool for learning GNNs.
Through a review of existing GNN educational materials and close collaboration with experts in teaching GNNs, we designed and implemented four main features to facilitate the learning process: hierarchical level of details, directional math-visualization linking, complementary views, and exploration of data, models, and tasks.
We currently deploy \name\ in three GNN-related courses across three universities.
The results from this deployment, including reported usage scenarios from users in GNN-related courses, interviews with TAs and students, and user activities in the wild, show the usability and effectiveness of \name.
We believe that the proposed visualization techniques and the derived design lessons have implications beyond GNNs, inspiring future studies on visualization tools for AI education.



\bibliographystyle{IEEEtran}
\bibliography{template}

\vspace{-38pt}

\begin{IEEEbiography}[{\includegraphics[width=1in,height=1.25in,clip,keepaspectratio,trim=0 0 0 0]{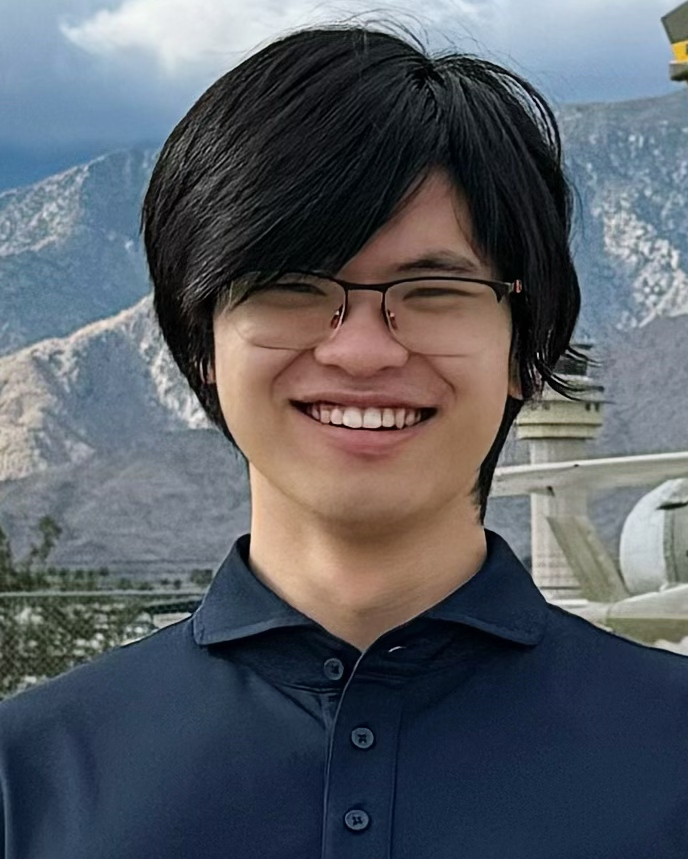}}]{Yilin Lu}
is currently working toward his Bachelor degree in computer science, mathematics, and economics with the College of Liberal Arts, University of Minnesota, Twin Cities. His research interests include explainable AI,  human-centered AI, and interactive visualization systems. More information can be found at \url{https://sites.google.com/umn.edu/yilinharry-lu/bio}. 
\end{IEEEbiography}

\vspace{-38pt}

\begin{IEEEbiography}[{\includegraphics[width=1in,height=1.25in,clip,keepaspectratio]{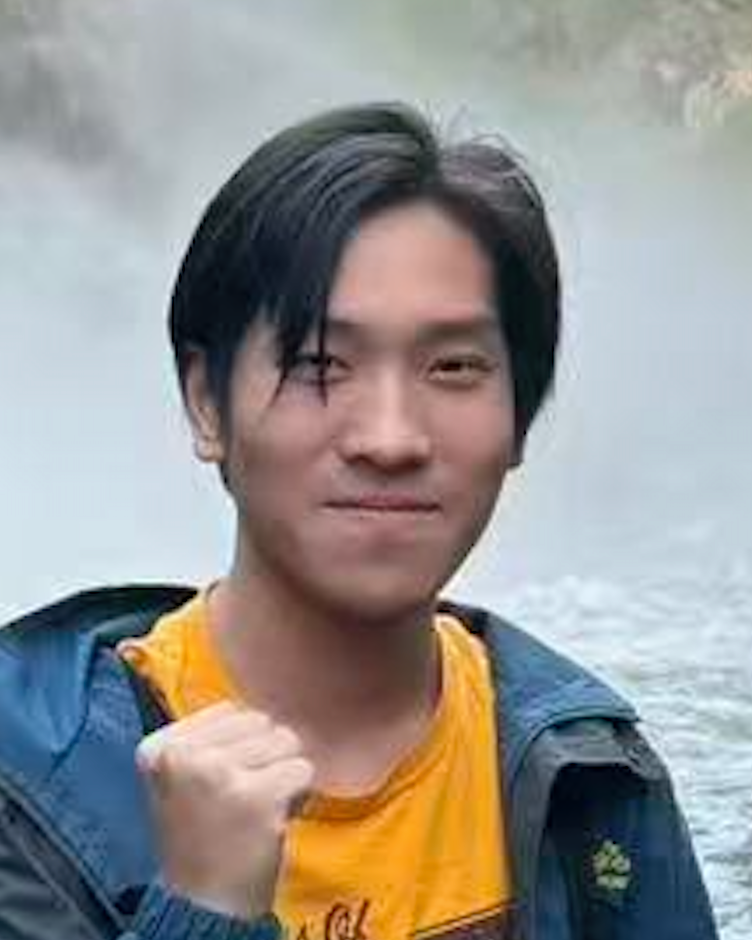}}]
{Chongwei Chen}
is currently an undergraduate student majoring in Mathematics at the University of Minnesota, Twin Cities. His research centered on optimization and machine learning. 
\end{IEEEbiography}

\vspace{-38pt}

\begin{IEEEbiography}[{\includegraphics[width=1in,height=1.25in,clip,keepaspectratio]{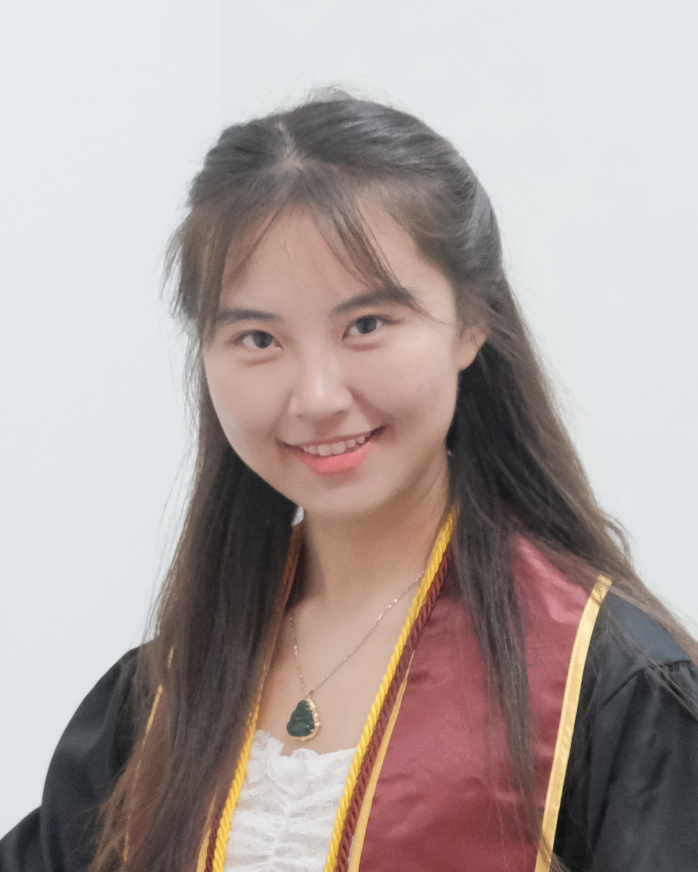}}]
{Yuxin Chen}
 is a Research Assistant in the Department of Computer Science and Engineering at the University of Minnesota, where she received the B.S. degree in computer science in 2025. Her research interests include human-centered natural language processing and interactive AI systems. She has worked on research projects analyzing AI–human collaborative writing patterns and developing visualization methods to improve reasoning interpretability. More information is available at \url{https://richsomeday222.github.io/}. 
\end{IEEEbiography}

\vspace{-35pt}

\begin{IEEEbiography}[{\includegraphics[width=1in,height=1.25in,clip,keepaspectratio]{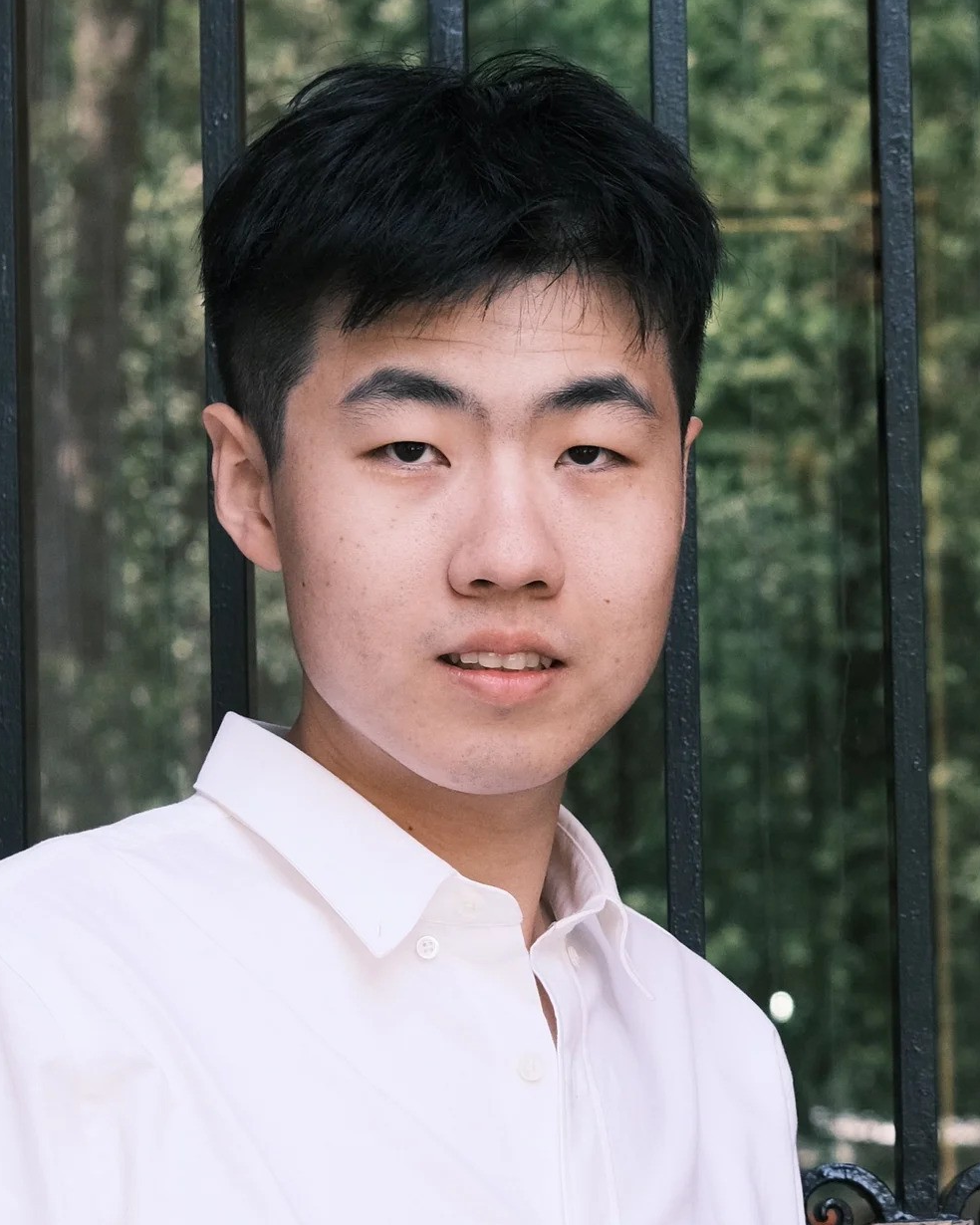}}]
{Kexin Huang}
  is a Ph.D. student in Computer Science at Stanford University, advised by Prof. Jure Leskovec and affiliated with the Stanford AI Lab, supported by the Stanford Bio-X Fellowship. His research focuses on developing interpretable, deployable AI systems for biomedical and therapeutic discovery, with an emphasis on multi-modal modeling, agentic reasoning, and uncertainty quantification.
\end{IEEEbiography}

\vspace{-38pt}

\begin{IEEEbiography}[{\includegraphics[width=1in,height=1.25in,clip,keepaspectratio]{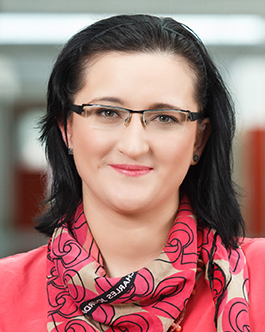}}]
{Marinka Zitnik}
 is an Associate Professor in the Department of Biomedical Informatics at Harvard Medical School, Associate Faculty at the Kempner Institute for the Study of Natural and Artificial Intelligence at Harvard University, Associate Member at the Broad Institute of MIT and Harvard, and Affiliated Faculty at the Harvard Data Science Initiative. 
\end{IEEEbiography}

\vspace{-38pt}

\begin{IEEEbiography}[{\includegraphics[width=1in,height=1.25in,clip,keepaspectratio]{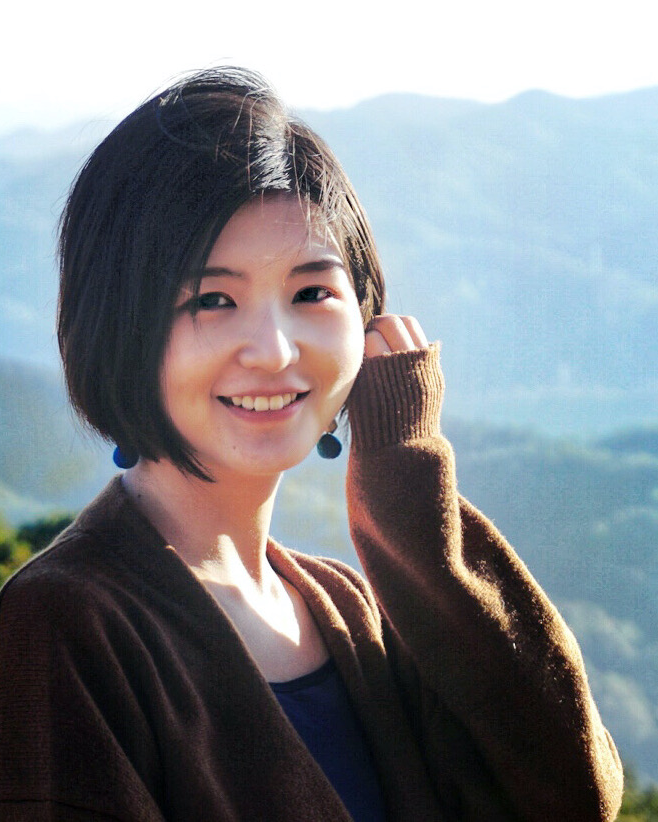}}]
{Qianwen Wang}
is an Assistant Professor in the department of computer science and engineering at the University of Minnesota. Her research focuses on improving communication and collaboration between domain experts and AI through the design and development of interactive visualization systems. Her studies aim to advance the understanding of human–AI interaction dynamics, support interpretable AI, and enable intuitive user feedback to guide AI systems.  She received her Ph.D from Hong Kong University of Science and Technology. More information is available at \url{https://qianwen.info}.
\end{IEEEbiography}


\vfill

\end{document}